Dec 21, 2005

# Advances in Thick GEM-like gaseous electron multipliers

# Part II: low-pressure operation

**C.K. Shalem, R. Chechik[1], A. Breskin, K. Michaeli and N. Ben-Haim**

*Dept. of Particle Physics,*

*The Weizmann Institute of Science, 76100 Rehovot, Israel.*

**Abstract**

We present results on THGEM operation at low gas pressures, in the range 0.5-50 Torr. Gaseous gain up to $10^5$ was recorded at 10 Torr of Isobutane in a single multiplier element, with pulses rise-time in the few ns range. Cascaded operation of two THGEMs is reported, with similar total gain but at lower voltages on each element. The present work encompasses a large variety of THGEM geometries and includes electric-field distribution calculations and electron transport simulations. We discuss the operation mechanism and some possible applications of THGEMs at low gas pressures.

[1] Corresponding author. Email: Rachel.Chechik@Weizmann.ac.il; Fax:+97289242611

## 1. Introduction

The THGEM presented in the previous, related article [1], is based on avalanche multiplication in independent holes, arranged in a compact array. The holes, of 0.3 - 2 mm diameter, are mechanically drilled in a Cu-clad double-sided 0.4 – 3.2 mm thick G-10 plate, and then the Cu is etched to produce a 0.1 mm clearance from the hole rim. Radiation-induced electrons, deposited in the gas gap above the THGEM or produced in a solid converter placed above or deposited directly on the THGEM top face, drift towards the THGEM holes. The strong dipole electric field established in the holes by the potential difference between the two THGEM faces, focuses the electrons into the holes, whereby they are multiplied in the strong $E_{hole}$ field. An extraction field in the gap below the THGEM is responsible for the charge collection onto a readout anode or another multiplying electrode; a reversed extraction field will divert all charges to the THGEM bottom. The THGEM's operation principle is similar to that of the standard GEM [2], with the main advantages of suppressed photon-mediated secondary processes and true pixilated operation; the large hole dimensions result in electrode robustness, better electron focusing into the holes and higher total gains but coarser localization and somewhat slower pulses.

The previous article [1] presented our systematic study of the properties of these devices operated with various gases at atmospheric pressure; therefore it concentrated mostly on THGEMs of a geometry that is the most appropriate for atmospheric pressure applications, e.g. having a thickness of 0.4 mm. A previous study [3] described in some detail results obtained with a 1.6 mm thick electrode operated at atmospheric and at low gas pressures.

An important range of applications exists at low gas pressures, down to the sub-Torr values, for which the THGEM might be a very appropriate solution. Examples are in detection of heavily ionizing radiation with very low penetration, e.g. heavy particles in nuclear reactions. We are currently investigating the application of THGEMs for the readout of a low pressure (~100Torr) TPC operating in an Oxygen-rich gas mixture, for the study of the O($\gamma$,$\alpha$)C reaction in cosmologically-relevant energy range [4]. Another example is in the detection of very low-energy ions, recently proposed by us for track-structure studies in nanodosimetry applications. In this case the ions produced by the incident radiation in a model gas volume carry the information on the primary ionizations and their spatial correlations. For this purpose an imaging ion detector has to be developed that operates at sub-Torr gas pressure; the THGEM electron multiplier could become an important component of this ion detector.

Nuclear physics experiments have been successfully applying traditional multiwire proportional chambers (MWPCs) and parallel-plate avalanche chambers (PPACs) for the fast and efficient detection of heavily ionizing light and massive particles, at low gas pressures down to 1 Torr and below [5-9]. It was shown that the high gains, of ~$10^6$ or more, reached at very low gas pressures, make MWPCs and multistep avalanche chambers suitable tools for the detection and imaging of single electrons [10] and single photons [11]. In this article we present the results obtained with various THGEM electrodes operated in low-pressure gases, and discuss their properties. This article is preceded by Part I [1], dedicated to the general description of the THGEM operation, methodology and properties at atmospheric pressure.

## 2. Methodology

Like our studies at atmospheric pressure, the methodology encompassed electric-field calculations (MAXWELL) [12], electron-transport and avalanche simulations (GARFIELD) [13] and experimental investigations. The electrode production and the investigated geometries as well the detailed description of the methodology are provided in [1]. It should be noted that unlike atmospheric pressure, GARFIELD electron-transport simulations at the low-pressure range did not match quantitatively the experimental data; the reason is yet unknown to us.

Our low-pressure operation studies included THGEM electrodes coupled to semitransparent photocathodes (*ST*PC) placed above the THGEM or reflective ones (*Ref*PC) deposited on the top face of the THGEM. We deposited CsI photocathodes of 30 or 300nm thickness for the *ST*PC and the *Ref*PC, respectively. The first was deposited on a quartz window pre-coated with a thin (3nm) Cr contact layer. We employed continuous or pulsed UV light sources, and recorded currents from various electrodes with the help of a precision electrometer (KIETHLY 610C). Individual power supplies were used on each electrode, to allow maximum powering flexibility. Multipliers comprising single- and double-THGEM elements were investigated; in the latter configuration the electrodes were mounted at least 6-10 mm apart. More details are provided in [1].

The experiments included mainly absolute effective gain (AEG) measurements with single and double-THGEMs. It should be noted that the gain is measured by recording currents at the bottom electrode of the THGEM and not on a readout anode as common in the literature. In a single THGEM the AEG is defined as the product of the efficiency to transport the primary electron from its creation point into the holes (electron transfer efficiency, ETE) and the absolute gaseous multiplication in the holes. In a double-THGEM cascade the AEG includes the absolute multiplication in both THGEMs and the efficiency to transport the electron from its creation point to the first THGEM holes (ETE) as well as the efficiency to extract it from the first THGEM into the holes of the second one. When gain permitted, we recorded single-photoelectron fast pulses and registered their rise time.

Figure 2 depicts the experimental scheme for AEG measurements, in the *ST* and *Ref* modes. It is done by DC current recording, in two steps: 1) the normalization step, providing the total photocurrent hitting the THGEM top face and 2) the AEG measurement, providing the total electrons current collected at the THGEM bottom with a slightly reversed $E_{trans}$ field below it. In the AEG step the field $E_{drift}$ above the THGEM was kept 0 in the *Ref* mode and typically 30 V/cm·Torr in the *ST* mode. The ratio of the two currents provides the AEG. A similar scheme and procedure were used with double THGEMs: the normalization step was identical; the AEG step was performed with $E_{trans}$ between the two elements set typically to ~20V/ cm·Torr.

## 3. Results and discussion

The reader is referred to table I in the related paper [1], which summarizes the geometrical parameters of all THGEM electrodes used in this study. Their numbers in the following text refer to this table.

### *3.1 Electric-field calculations*

Figure 3 shows $E_{hole}$, the electric filed along the hole axis, calculated by MAXWELL for two different THGEMs for their respective optimal operation conditions at atmospheric ($Ar/CO_2$(70:30)) and at low pressure (10 Torr Isobutane). The THGEMs are those found experimentally to yield the highest gain in the respective pressures: THGEM#9 with t=0.4 mm and 0.3 diameter holes is suitable for atmospheric pressure operation while THGEM#4 or THGEM#6 with t=1.6mm and 1mm diameter holes are more appropriate for the p<50Torr range. The reader is referred to [1] for a more extended discussion on the electric-field systematics: e.g. the amount of field confinement as function of hole diameter (for a fixed plate thickness) and the effect of an external field (e.g. $E_{drift}$ or $E_{trans}$) on the profile of $E_{hole}$. With the thicker plates and larger hole diameter employed at low pressure, the avalanche spread outside the holes is expected to be more pronounced.

### *3.2 Absolute effective gain*

Figures 4-9,12 summarize the results of absolute effective gain (AEG) measured at low-pressure Isobutane and Ar/$CO_2$ (70:30). Six different THGEM geometries were investigated in 0.5,1,5,10, 30 and 50 Torr. In general, the THGEMs were found to be very robust under low-pressure operation, and not a single THGEM was damaged during our extensive measurements.

Figure 4 compares the AEGs of different THGEMs in 10 Torr Isobutane. THGEMs #1 and #2 were produced by drilling the PCB plates and then dipping them in acid to smooth-out sharp edges in the Cu, without etching a clearance in the copper layer around the hole rims. The additional etching stage of a 0.1mm clearance around the hole rim (added later to the production process), clearly improved the maximal gain by up to an order of magnitude (compare # 2 and #3). THGEM#11 yielded the highest gain in this pressure, and no advantage was gained by increasing the thickness from 2.2 to 3.2 mm. Compared to the standard GEM, also shown in the figure, the THGEMs provide AEG of a few orders of magnitude higher.

Figures 5 and 6 show AEG data measured at 1 and 0.5 Torr Isobutane, with different THGEMs. From Fig. 5 it is seen that all electrodes reach about the same maximal gain at 1 Torr; from Fig. 6 it is seen that THGEM#1 (without the etched clearance around the hole) yields an order of magnitude lower gain compared to the other electrodes. Nevertheless, the maximal AEG depends primarily on the gas pressure. This is evident from figures 7 and 8, comparing the gain of THGEMs #4 and #5, over the whole pressure range.

Figure 9 compares the AEG measured with single- and double-THGEM#11 in Isobutane, in the pressure range of 0.5 – 10 Torr. It is noted that about the same maximal gain is obtained with the double-THGEM, but at a lower operation voltage on each THGEM; it is interesting to note that at the lowest pressure of 0.5 Torr, the single-THGEM curve deviates from proportionality already at an effective gain of 100 but with double THGEM the curve reveals proportional operation up to the maximal effective gain of $10^4$. Double-THGEM is thus more proportional and more stable than single THGEM in this pressure range. When cascading two THGEMs at the low gas pressure conditions, it was found necessary to maintain a distance of the order of 6-10mm between the two elements; smaller distances resulted in very unstable operation and sparks. This could be understood in terms of the avalanche extending out of the holes at high multiplication conditions and the overlap of the fields from the two THGEMs.

Figures 10 and 11 show some AEG curves plotted on a linear scale; it accentuates the fact that the multiplications curves are parallel but there is a voltage shift of the multiplication onset as function of the THGEM geometry. It therefore also accentuates the dependence of gain uniformity on the THGEM fabrication precision.

Figure 12 depicts the AEG measured with THGEM#4 in low pressure Ar/$CO_2$(70:30). Compared to Isobutane, the operation voltages and the maximal attainable gains in this non-flammable gas are lower.

The high gain obtained with the THGEMs permitted recording single-photoelectron signals with a fast (1ns rise-time) current preamplifier. Some signals are shown on figure 13; a) a single-photoelectron pulse, recorded with THGEM#6 and a continuous Hg(Ar) lamp at 10 Torr Isobutane, and b) a multi-photoelectron pulse, recorded with THGEM#5 and a pulsed $H_2$ lamp at 1 Torr Isobutane. Both pulses show similar rise times, of 5 – 6 ns; they are due to the high drift velocity of the electrons and ions in the high reduced-electric fields (E/p) in these operation conditions.

## 4. Summary

Within this work, various THGEMs with hole diameter of 1mm, thickness of 1.6 – 3.2 mm and hole distance of 1.5 – 7 mm were tested with Isobutane and Ar/$CO_2$ (70:30), at low gas pressure in the range 0.5 – 50 Torr. Spark-free high effective gains, of $10^5$ and $10^7$, could be reached at 1 and 10 Torr Isobutane, respectively; the signals have rise times of typically 5-6 ns. All the THGEMs tested within this work showed very high robustness, and none was irreversibly damaged during operation. Spark-free double-THGEM operation was obtained when setting the transfer gap between the two elements to ~6-10mm, guaranteeing sufficient decoupling of the two avalanche processes; this wide gap is required probably because of the electric filed and avalanche extension outside the hole, by about the hole diameter, at high multiplication. The total effective gain in double-THGEM mode is similar to that obtained with a single-THGEM, but is achieved at lower operation voltage; this results in lower spark probability and better proportionality, particularly at the very low gas pressures (e.g. 0.5 Torr).

Large-area THGEMs of almost any geometrical shape can be produced with economical standard industrial process, over a large range of geometrical parameters and may find many applications in radiation detection and imaging. In particular, radiation detection in the low-pressure range may benefit from these robust devices. Examples of applications exist in nuclear physics, at accelerators or in space, where low-energy (~MeV) ions produced in vacuum should be efficiently detected; their low penetrability allows only very thin dead-layers at the entrance of the detectors, as for example thin windows that can withstand only very low gas pressures. Other examples are the detection of ions for various environmental or atmospheric-quality monitoring. In all cases the robust low-pressure THGEM might successfully replace the more fragile wire-chambers, though with some compromise on timing and localization precision.

**Acknowledgment**

This work was supported in part by the Benoziyo Center for High Energy Research, the Israel Science Foundation project No151-01 and by the Binational Science Foundation project No 2002240. C. S. was supported in part by the Fund for Victims of Terror of the Jewish Agency for Israel. A.B. is the W.P. Reuther Professor of Research in the peaceful use of Atomic Energy.

**Figure captions**

Figure 1. A microscope photograph (left) of a THGEM (right) with thickness t = 1.6 mm, hole diameter d= 1 mm and pitch a= 1.5 mm. A rim of 0.1 mm is etched around the mechanically drilled holes.

Figure 2. The effective-gain measurement setup with a single THGEM. 1) Normalization current measurement; 2) Multiplication current measurement. Left: using a semitransparent (ST) photocathode. Right: using a reflective (REF) photocathode deposited on the THGEM top surface.

Figure 3. The electric filed strength $E_{hole}$, along the hole axis, calculated by MAXWELL for two different THGEMs at optimal voltages for atmospheric $Ar/CO_2$(70:30) and 10 Torr Isobutane. At 10 Torr $E_{hole}$ is 10 times weaker but the reduced field $E_{hole}/p$ is 10 times higher. In both cases the field extends outside of the holes in proportion to the hole diameter (1mm for THGEM#4 and 0.3mm for THGEM#9).

Figure 4. Absolute effective gain measured with a semitransparent photocathode coupled to single THGEMs of different geometries numbered as in Table I of ref. [1], at 10 Torr Isobutane. Results with a standard GEM are shown for comparison.

Figure 5. Absolute effective gain measured with a semitransparent photocathode coupled to single THGEMs of different geometries, numbered as in Table 1 in ref. [1], at 1 Torr Isobutane.

Figure 6. Absolute effective gain measured with a semitransparent photocathode coupled to single THGEMs of different geometries, numbered as in Table 1 of ref. [1], at 0.5 Torr Isobutane. In this pressure range the multiplication process deviates from proportionality already at a gain of a few 100.

Figure 7. Absolute effective gain measured with a semitransparent photocathode coupled to a THGEM#4 (Table 1 in ref. [1]) in 1-50 Torr of Isobutane. This THGEM geometry yields the highest gains at 10-20 Torr.

Figure 8. Absolute effective gain measured with a semitransparent photocathode coupled to a THGEM#3 (Table 1 in ref. [1]), in 0.5 – 10 Torr of Isobutane.

Figure 9. The absolute effective gain measured with a semitransparent photocathode coupled to a THGEM#11 (Table 1 in ref. [1]) in single- and double-THGEM modes, in 0.5 - 10 Torr Isobutane (marked in the figure). $E_{drift}$ ~15V/cm·Torr in all cases; in the double-THGEM mode $E_{trans}$ between the two elements = 20, 7 and ~3V/cm·Torr for 10, 1 and 0.5 Torr, respectively. The maximal gain with two THGEMs in cascade is similar to that of a single one, but obtained at lower voltage on each element. At 0.5 Torr the multiplication process deviates from proportionality at single-THGEM gain of a few hundreds.

Figure 10. A difference of a few tens of Volts in the onset on the effective gain curve is evidently correlated with some differences in the geometrical parameters of the THGEM electrode such as the presence of an etched rim or the pitch size.

Figure 11. A difference of a few tens of Volts in the onset on the effective gain curve is evidently correlated with differences in the geometrical parameters of the THGEM electrode such as the thickness or the pitch size.

Figure 12. Absolute effective gain measured with semitransparent photocathode coupled to THGEM#4 (Table 1 in ref. [1]) in Ar/$CO_2$ (70:30) at different pressures.

Figure 13a) A fast single-photon pulse of ~5ns rise-time, measured with a continuous Hg(Ar) lamp in THGEM#6; p=10 Torr Isobutane, effective gain~$6\times10^5$. The after-pulses are due to electronic noise; b) A fast multi-photon pulse of ~6ns rise-time, measured with a pulsed $H_2$ lamp in THGEM#5; p=1 Torr Isobutane, effective gain $>10^4$.

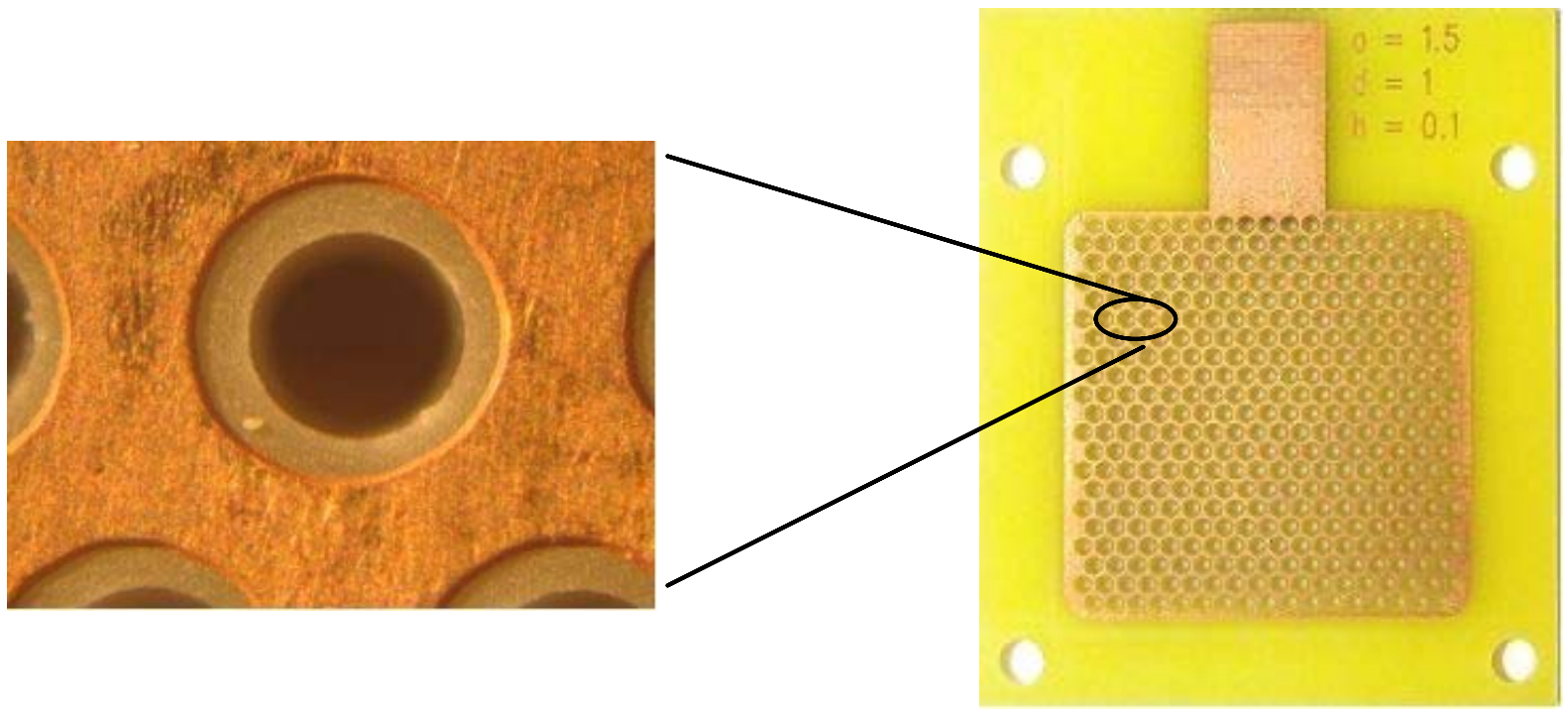

*Figure 1.*

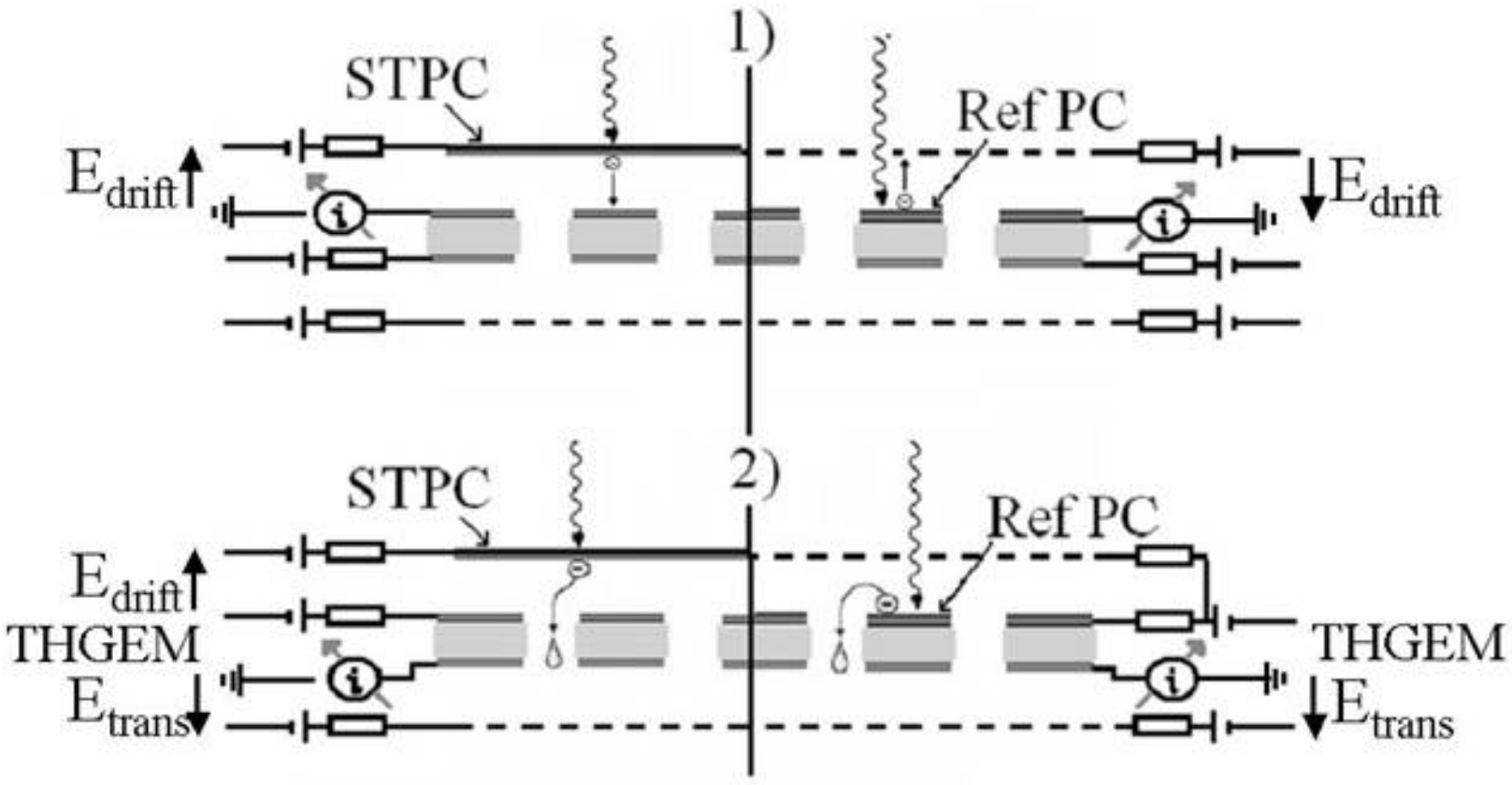

*Figure 2.*

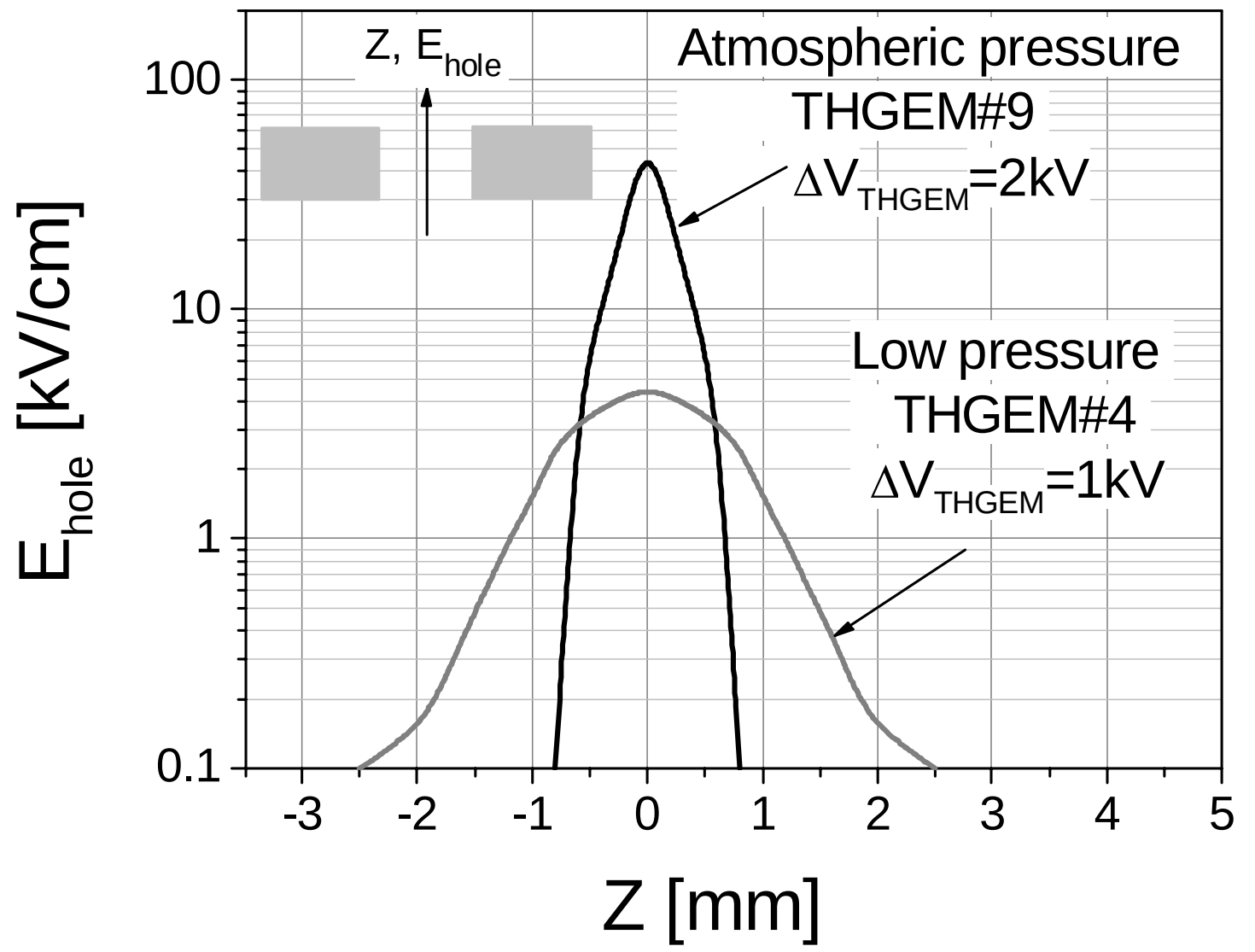

*Figure 3.*

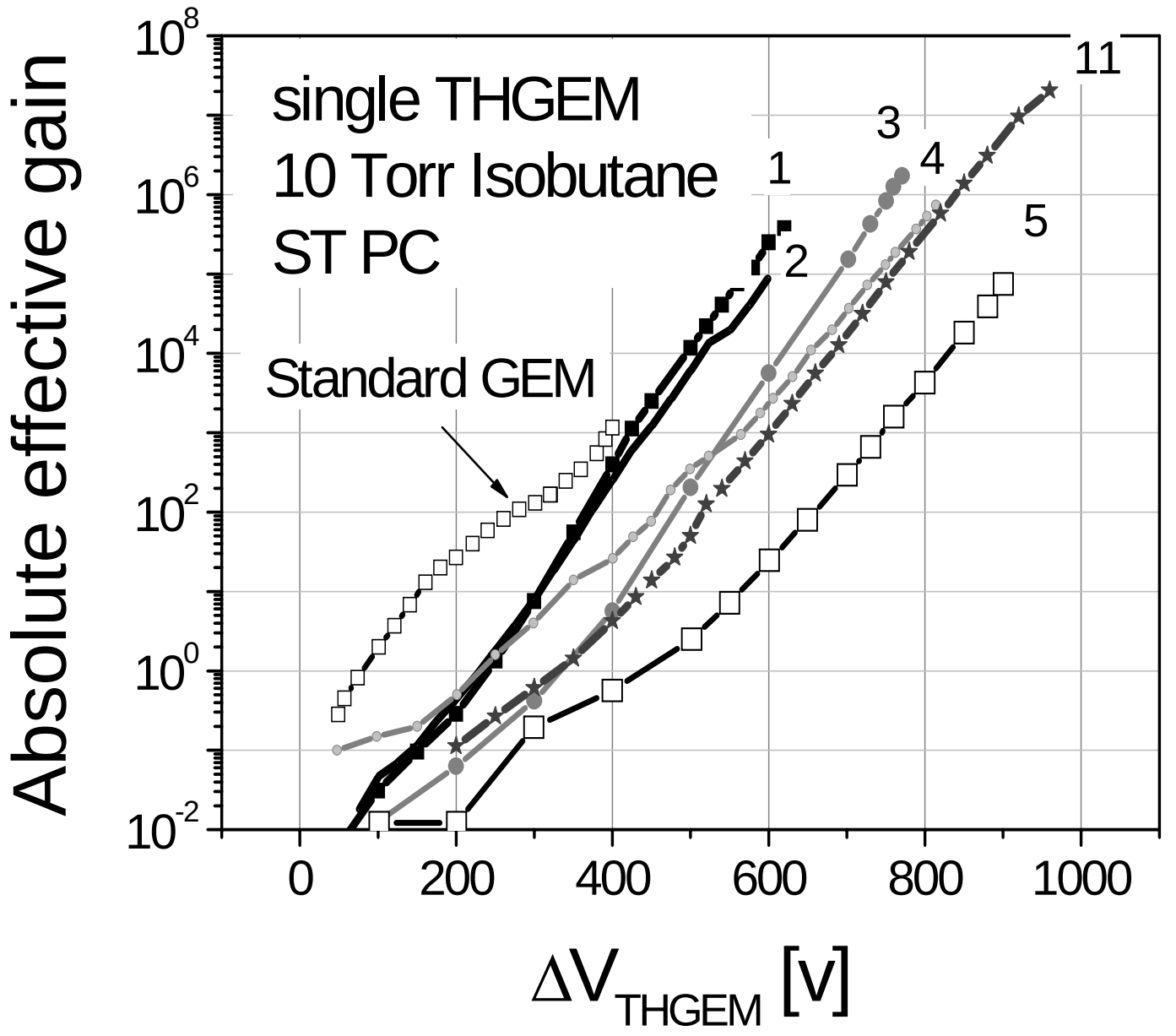

*Figure 4.*

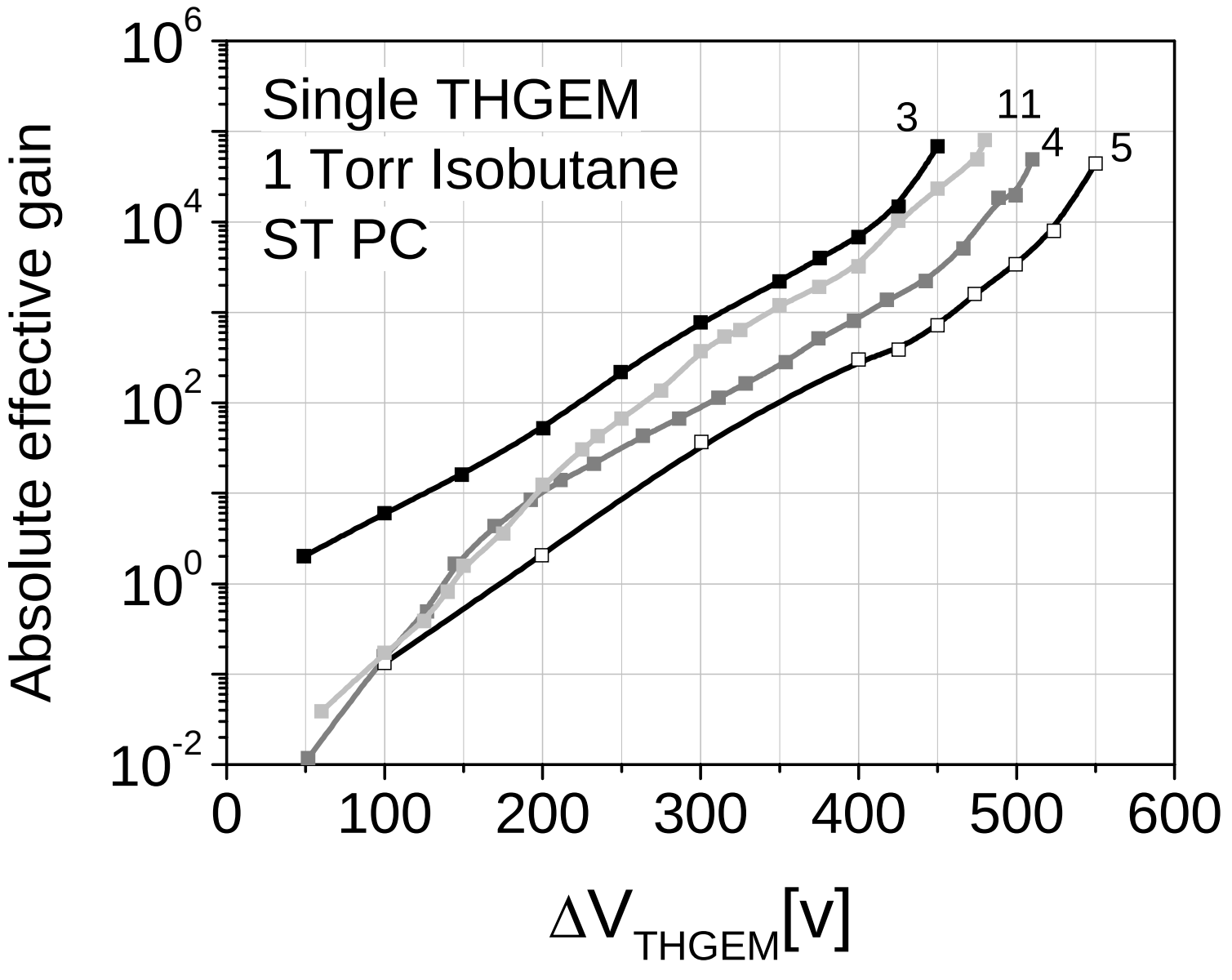

*Figure 5.*

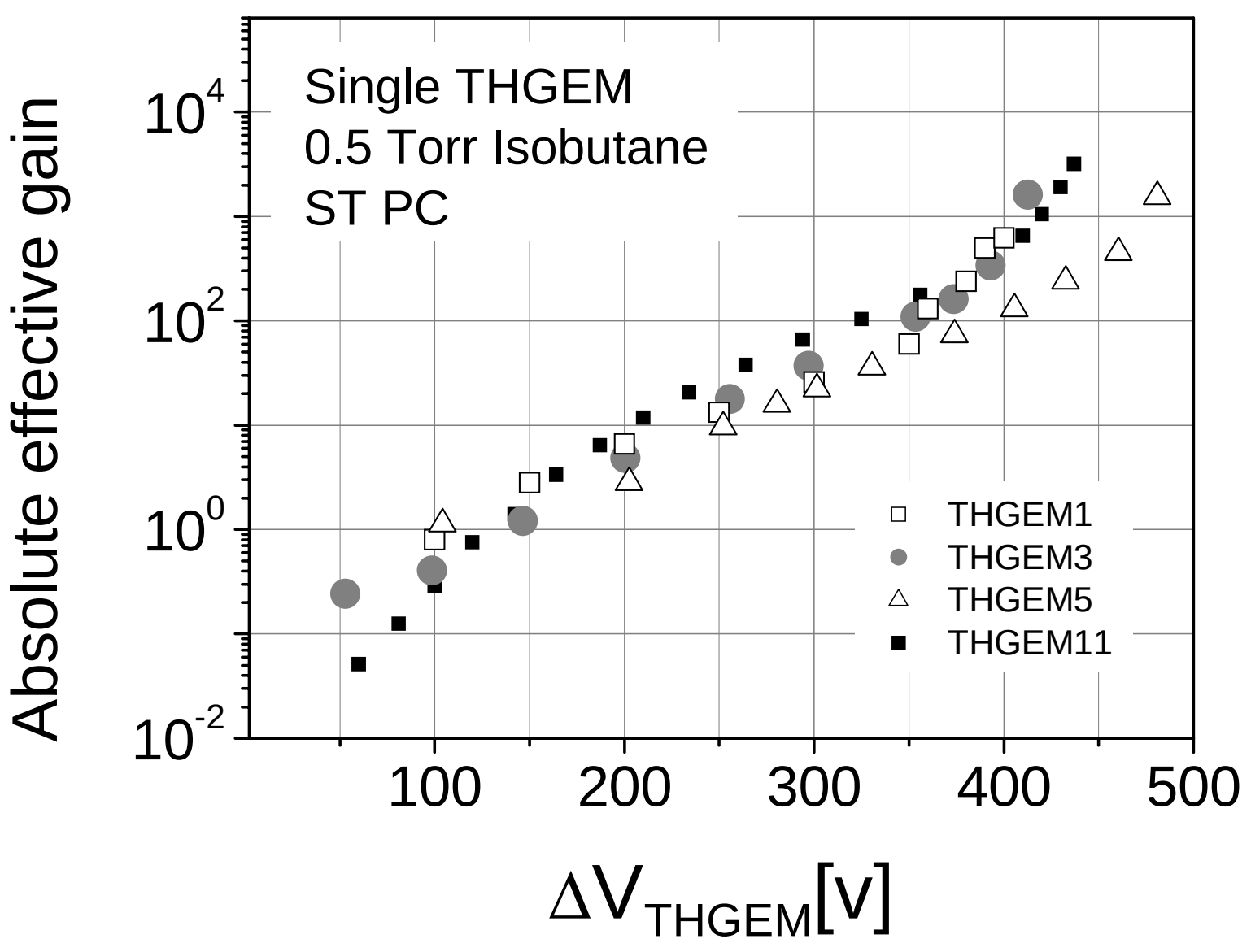

*Figure 6.*

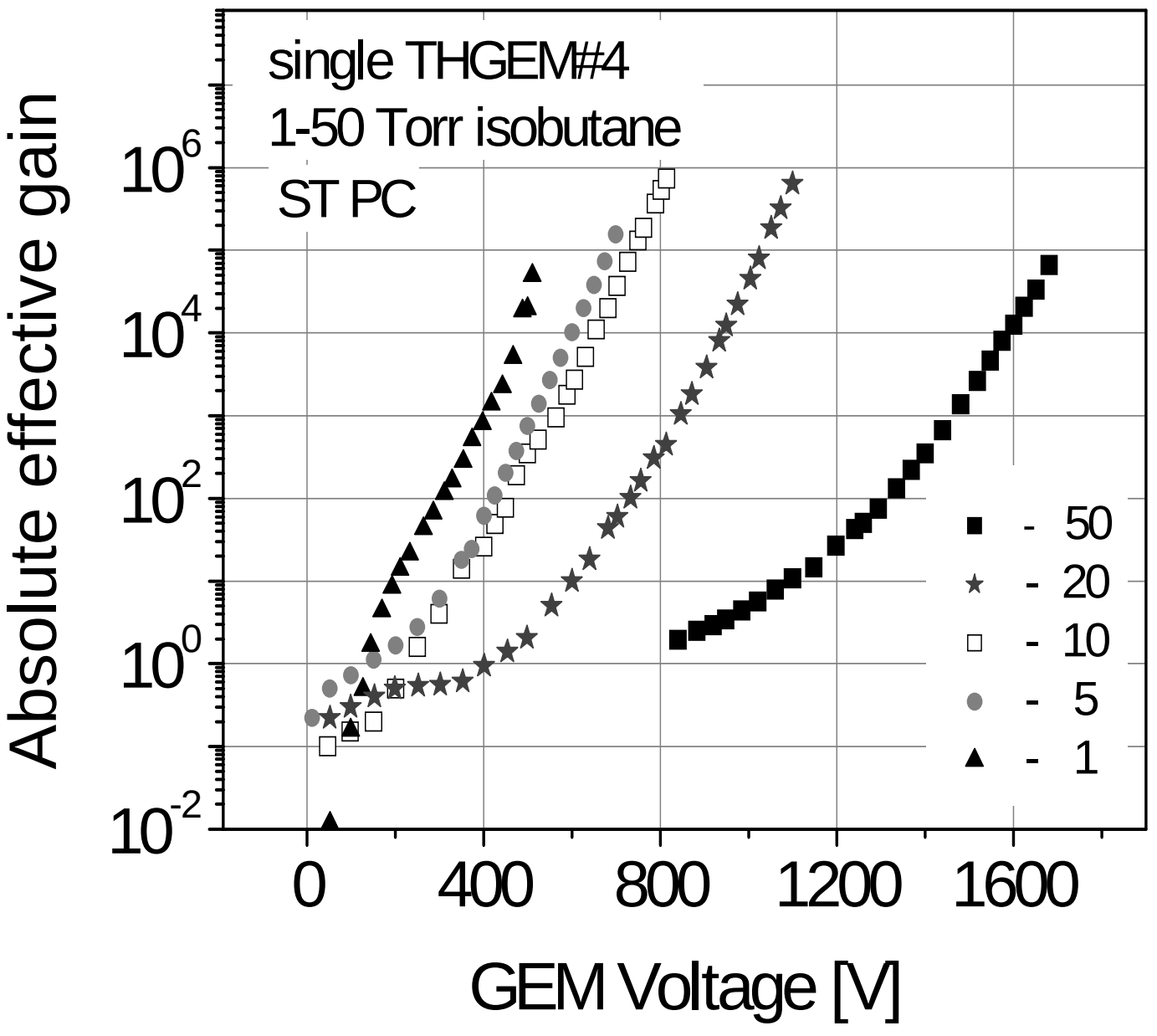

*Figure 7.*

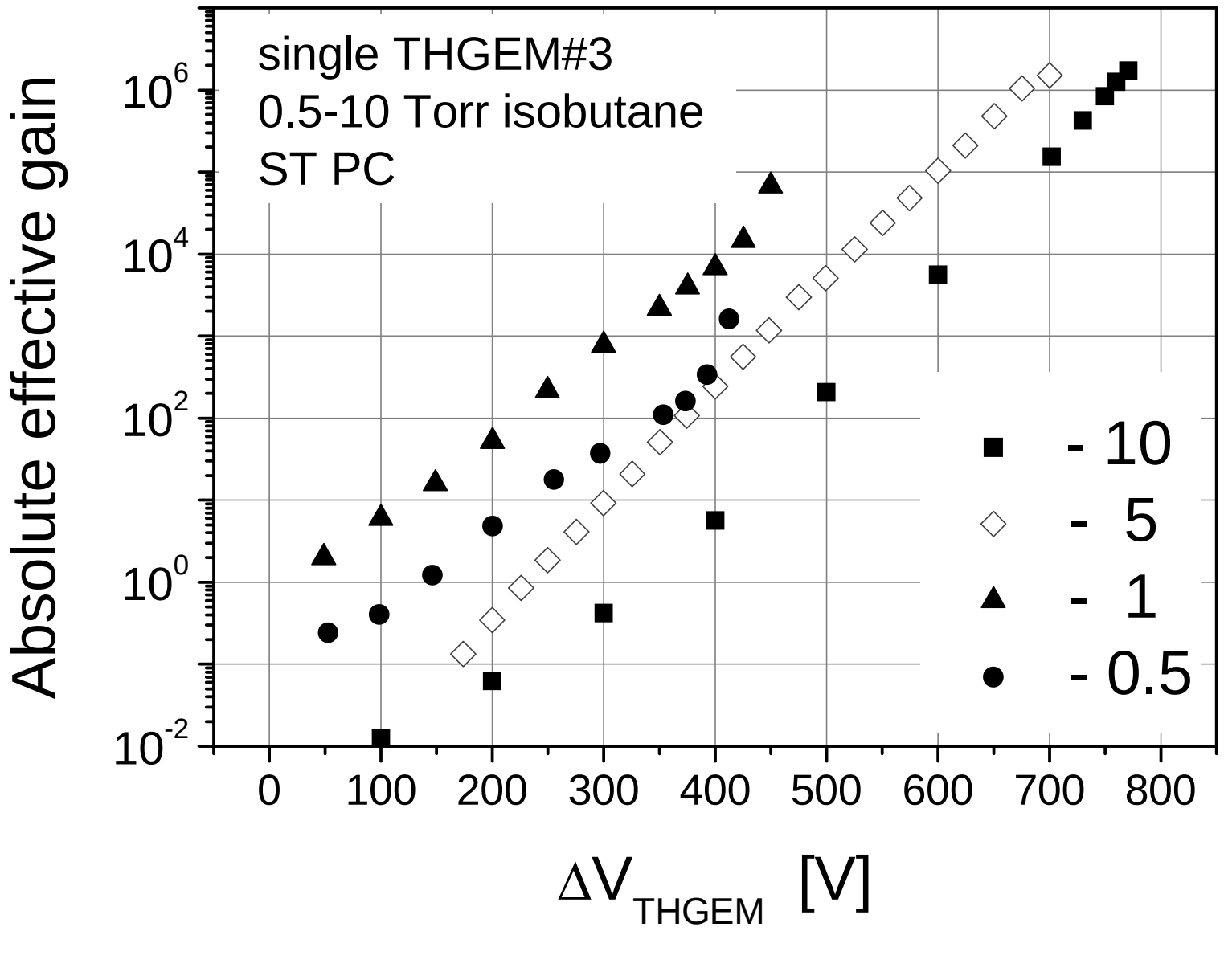

*Figure 8.*

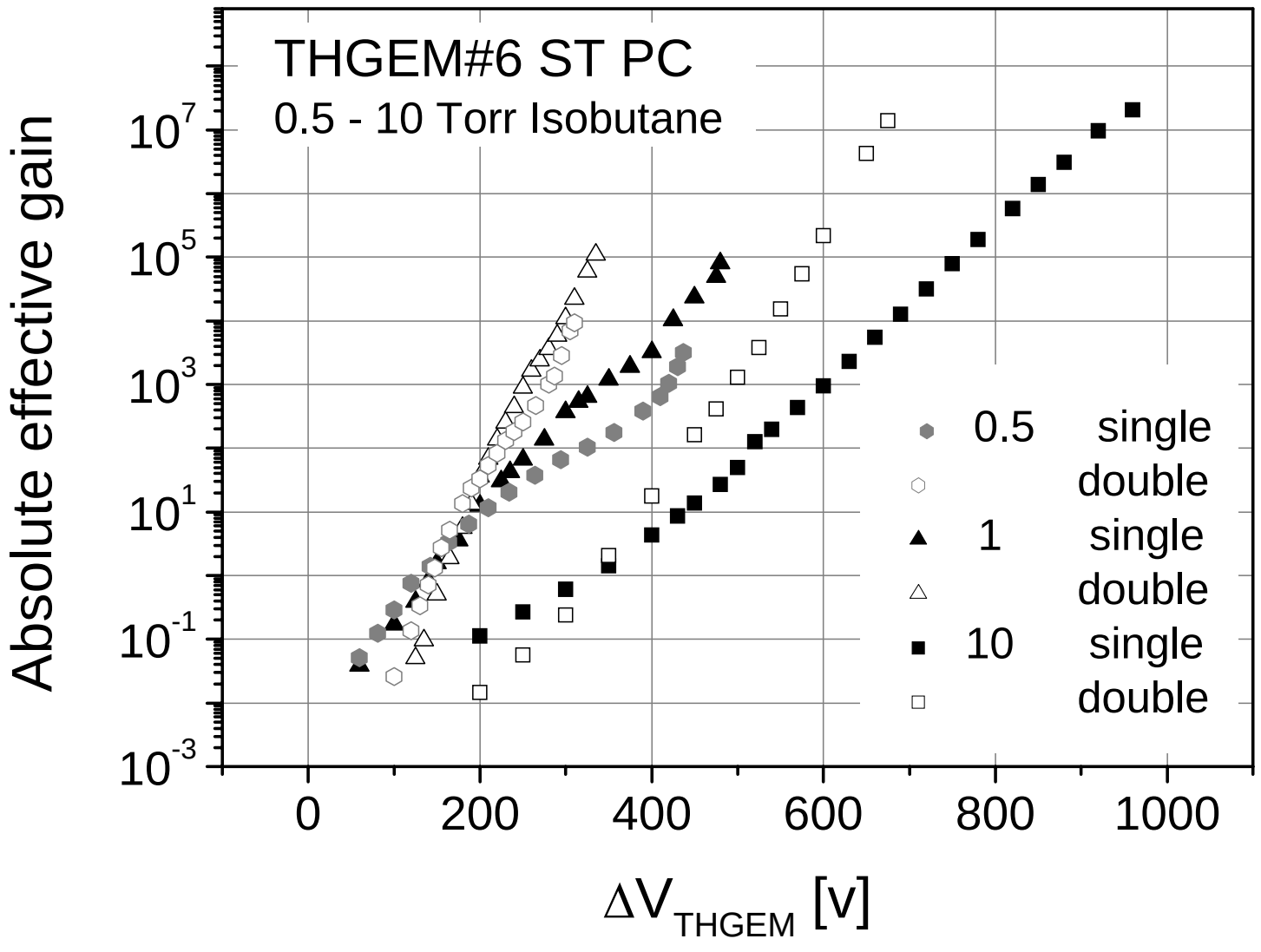

*Figure 9.*

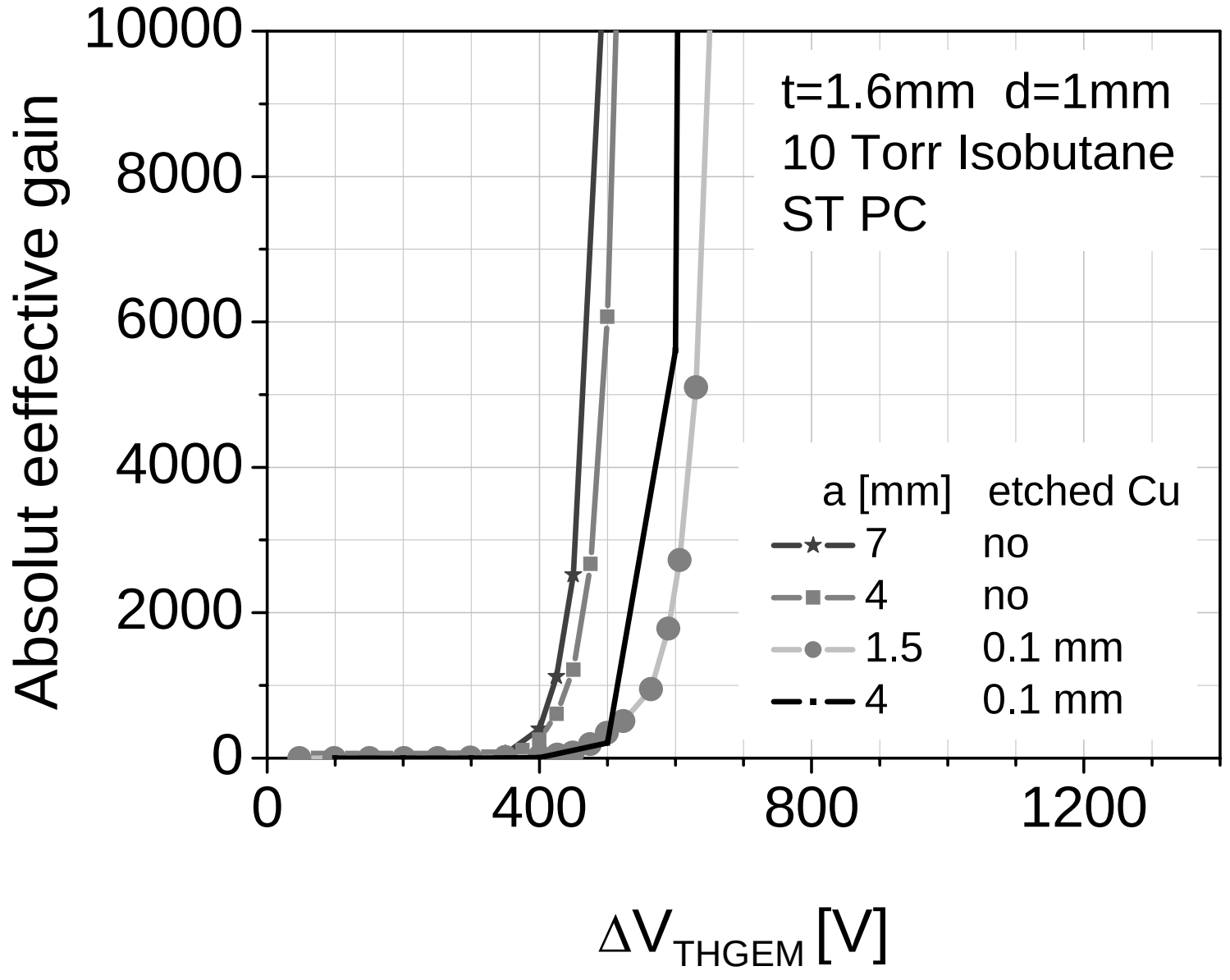

Figure 10.

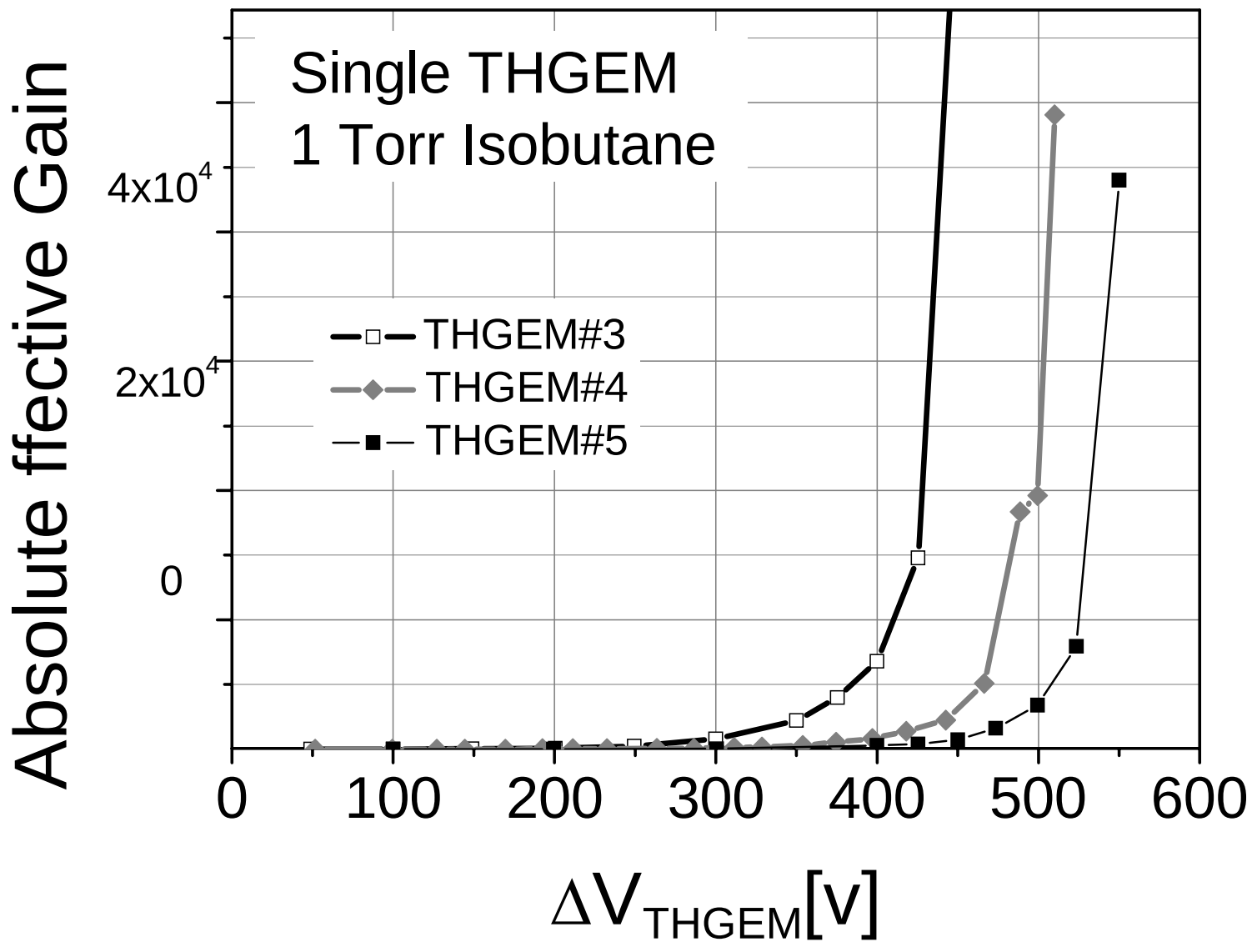

Figure 11.

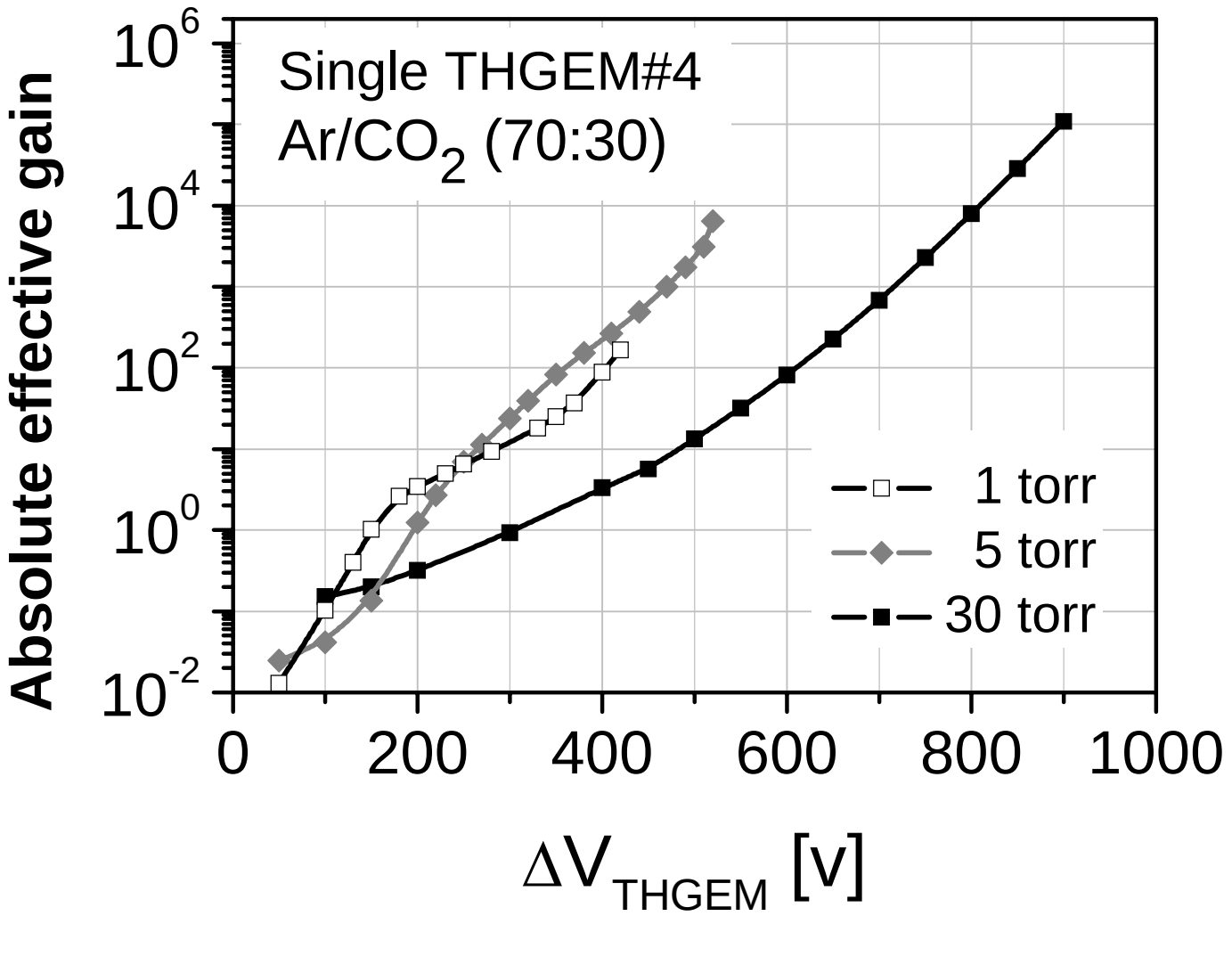

*Figure 12.*

a)

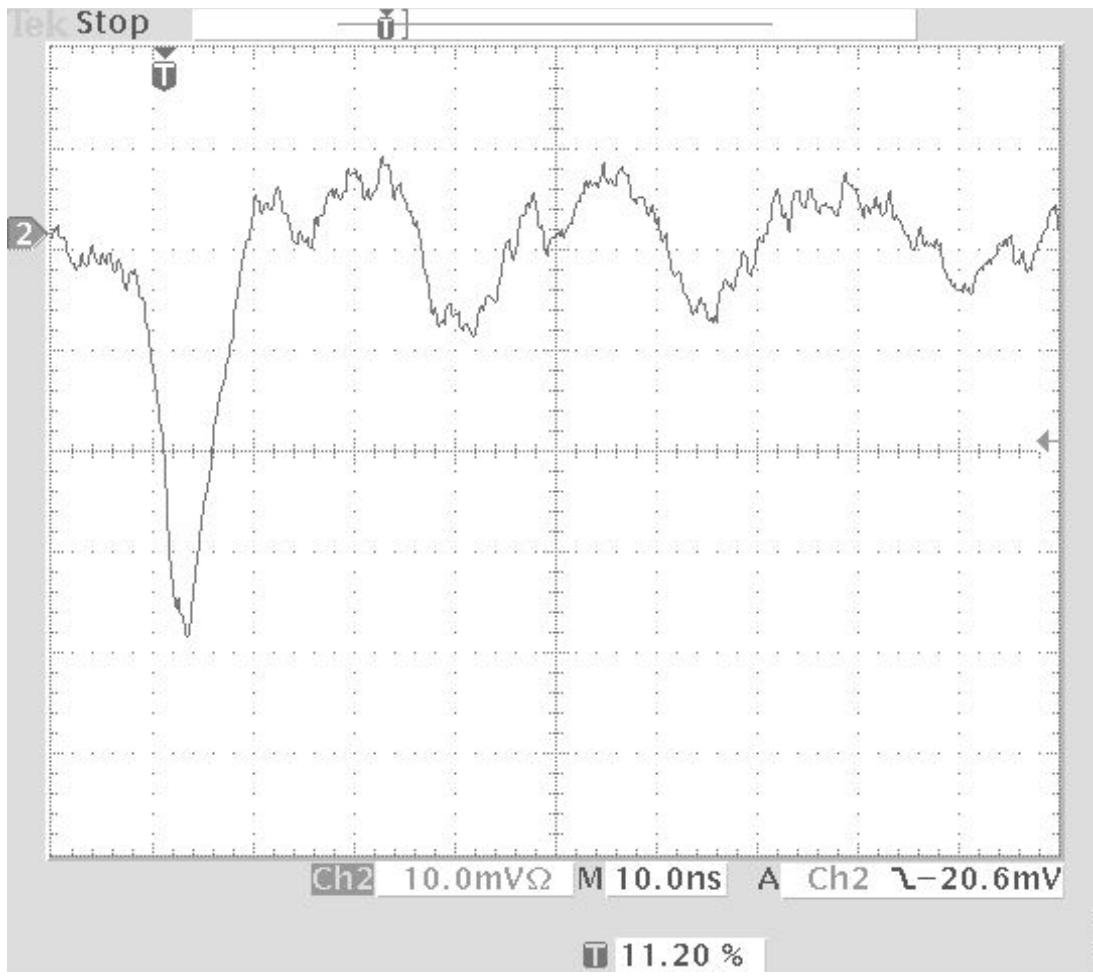
Tek Stop
Ch2 10.0mVΩ M 10.0ns A Ch2 ↘ −20.6mV
11.20 %

b)

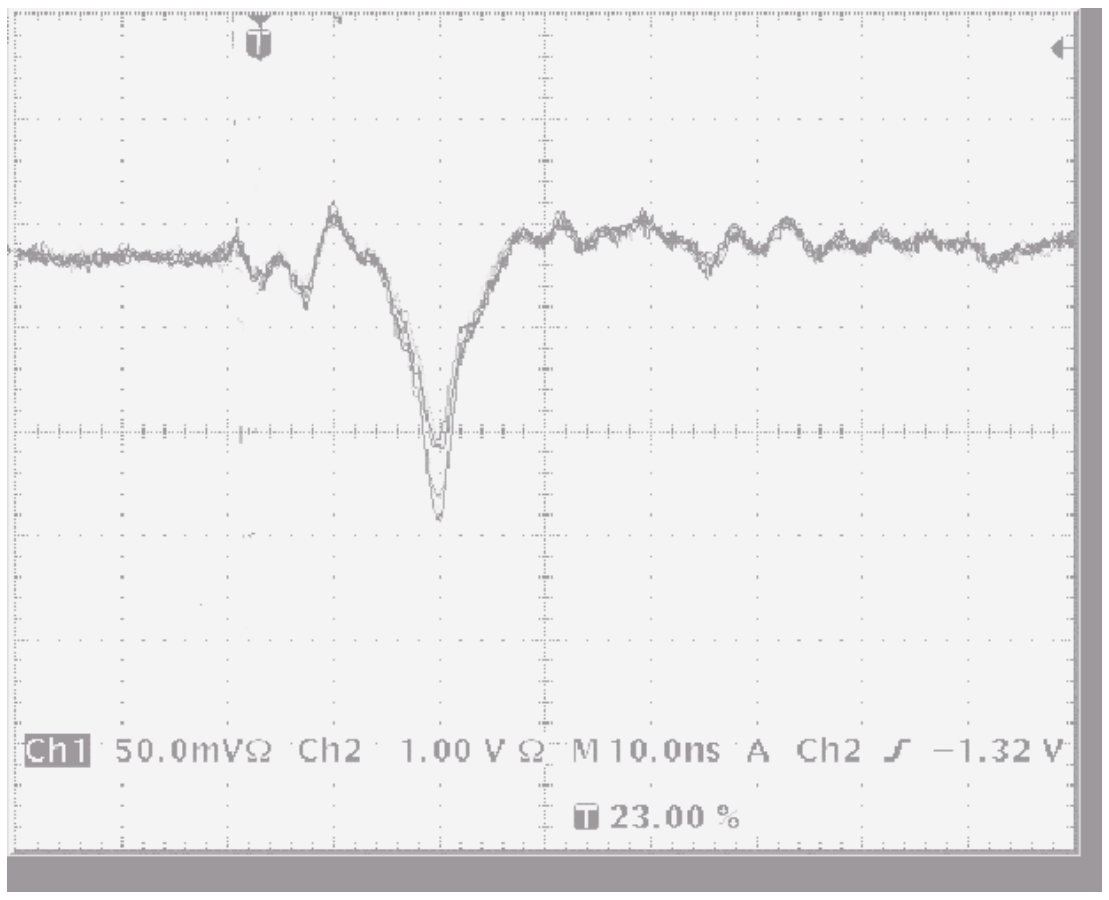
Ch1 50.0mVΩ Ch2 1.00 V Ω M 10.0ns A Ch2 ↗ −1.32 V
23.00 %

*Figure 13.*